\documentclass[pra,aps,12pt,showpacs,showkeys]{revtex4}
\usepackage{epsfig}

\begin{document}

\textbf{Fu and Chen Reply}: In the proceeding comment \cite{cc}, Sj\"{o}%
qvist states that the concept of mixed state geometric phase in
cyclic evolution suggested in our recent work \cite{fug} is not
gauge invariant.

To justify our concept and clarify the issue involved, we have to
review the concept of geometric phase. In general, it means that
the geometric phase is invariant under $U(1)$ transformation to
say the geometric phase is gauge invariant, i.e., the geometric
phase is a U(1) gauge invariant. The non-Abelian gauge geometric
phase has been suggested only if a set of quantum states remains
degenerate as the Hamiltonian varies \cite{zee}. The definition of
geometric phase of mixed states suggested in our recent work
\cite{fug} is a $U(1)$ gauge invariant.

Supposing a quantum system with the Hamiltonian $H(t),$ the density operator
$\rho (t)$ of this system will undergo the following evolution
\begin{equation}
\rho (t)=U(t)\rho (0)U^{+}(t),  \label{evo}
\end{equation}
where $U(t)=\mathbf{T}e^{-i\int_{0}^{t}H(t^{\prime })dt^{\prime
}/\hbar }$, here $\mathbf{T}$ is the chronological operator. If
$[U(\tau ),\rho (0)]=0,$ i.e., $\rho (\tau )=\rho (0),$ we say
this state undergoes a cyclic evolution with period $\tau $. In
Ref. \cite{fug}, we suggested the geometric phase $\phi _{g}$ as
\begin{equation}
\phi _{g}=\phi -\phi _{d},  \label{all}
\end{equation}
where
\begin{equation}
\phi =\arg Tr[\rho (0)U(\tau )].  \label{tp}
\end{equation}
is the total phase, and
\begin{equation}
\phi _{d}=-i\int_{0}^{\tau }dtTr\left[ \rho (0)U^{+}(t)\frac{dU(t)}{dt}%
\right] ,  \nonumber
\end{equation}
is just the dynamical phase during the cyclic evolution.

The geometric phase can also be expressed as
\begin{equation}
\phi _{g}=\oint \beta ,  \label{gee2}
\end{equation}
and
\begin{equation}
\beta =iTr\left[ \rho (0)\widetilde{U}^{+}(t)d\widetilde{U}(t)\right] .
\label{g1}
\end{equation}
where $\widetilde{U}(t)=e^{-i\phi (t)}U(t)$ such that $\phi (\tau )=\phi .$ $%
\beta $ is a canonical one-form in the parameter space. The $U(1)$ invariant
property of the equation (\ref{gee2}) has been discussed in detail in Refs.
\cite{fug,pg,nonc2}.

If $U(\tau )=e^{i\phi }I$, i.e., $U(\tau )$ is a global cyclic
evolution, we proved in Ref. \cite{fug} that the geometric phase
can be expressed as
\begin{equation}
\phi _{g}=\phi -\sum\limits_{k}w_{k}\phi _{d}^{k}=\sum\limits_{k}w_{k}\phi
_{g}^{k}.  \label{sumg}
\end{equation}

However, if we take the transformation suggested in the comment
\cite{cc},
\begin{equation}
V(t)=\sum e^{-i\alpha _{k}(t)}\left| \psi _{k}\right\rangle \left\langle
\psi _{k}\right| ,  \label{uu}
\end{equation}
with $\alpha _{k}(\tau )-\alpha _{k}(0)=2\pi n_{k},$ then the corresponding
unitrary transformation is $U^{\prime }(\tau )\left| \psi _{k}\right\rangle
=U(\tau )V(\tau )\left| \psi _{k}\right\rangle =e^{i(\phi +2\pi
n_{k})}\left| \psi _{k}\right\rangle .$ So, $U^{\prime }(\tau )$ is no more
a global cyclic evolution, hence the $\phi _{g}$ can not be expressed as $%
\sum\limits_{k}w_{k}\phi _{g}^{\prime k}$ unless $n_{k}$ is a constant for
any $k$.

In fact, (\ref{uu}) is not a $U(1)$ transformation when $n_{k}$ is
not a constant for any $k$. One does not need the definition of
geometric phase would be a gauge invariant under such a
non-Abelian transformation, since the geometric phase is only a
$U(1)$ gauge invariant in general.

In Ref. \cite{Ekert}, Sj\"{o}qvist \textit{et al}., proposed a
definition of geometric phase for mixed states under the parallel
transport condition. We have proven in our work \cite{fug} that
our definition consists with the definition of them when $U(t)$
satisfies the parallel transport condition. Using nuclear magnetic
resonance technique, Du \textit{et al}. have observed the
geometric phase when $U(t)$ satisfies the parallel transport
condition \cite {exp}. In Ref. \cite{fug}, we have shown that our
predications consist with what have been observed by Du and his
co-works. Unfortunately, Du and his co-works have not designed to
measure the geometric phase for general cases (for the cases that
$U(t)$ is not a parallel transport). In fact, the dynamic phase
can be eliminated by ``spin echo'' method if $U(t)$ is not a
parallel transport \cite{echo}, so the geometric phase suggested
by Eq. (\ref{all}) can be observed experimentally.

In summary, the geometric phase suggested in Ref. \cite{fug} is
$U(1)$ gauge invariant and can be observed by nowadays technique.
It is improper to demand the geometric phase as a non-Abelian
structure in general.


\begin{thebibliography}{9}
\bibitem{cc}  E. Sj\"{o}qvist, ''Comment on ''Geometric phases for mixed
states during cyclic evolutions.

\bibitem{fug}  L.-B. Fu and J.-L. Chen, J. Phys. A \textbf{37}, 3699 (2004).

\bibitem{zee}  F. Wilczek and A. Zee, Phys. Rev. Lett. \textbf{52}, 2111
(1984); Phys. Rev. A, \textbf{38}, 1 (1988).

\bibitem{pg}  D.N. Page, Phys. Rev. A \textbf{36}, 3479 (1987).

\bibitem{nonc2}  A. K. Pati, Phys. Rev. A \textbf{52}, 2576 (1995); A.K.
Pati, J. Phys. A: Maht. Gen. \textbf{28}, 2087 (1995).

\bibitem{Ekert}  E. Sj\"{o}qvist, A.K. Pati, A. Ekert, J.S. Anandan, M.
Ericsson, D.K.L. Oi, and V. Vedral, Phys. Rev. Lett. \textbf{85}, 2845
(2000).

\bibitem{exp}  J. Du, P. Zou, M. Shi, L.C. Kwek, J.-W, Pan, C.H. Oh, A.
Ekert, D.K.L. Oi, and M. Ericsson, Phys. Rev. Lett. \textbf{91}, 100403
(2003).

\bibitem{echo} R.A. Bertlmann, K. Durstberger, Y. Hasegawa, and B.C. Hiesmayr
Phys. Rev. A \textbf{69}, 032112 (2004).

\end{thebibliography}
\end{document}